\documentclass[runningheads]{llncs}
\usepackage[T1]{fontenc}
\usepackage{graphicx}
\usepackage{eso-pic}
\usepackage{tabularx}
\usepackage{comment}
\usepackage{url}
\usepackage[misc,geometry]{ifsym}
\begin{document}
\title{Security Analysis of Top-Ranked mHealth Fitness Apps: An Empirical Study}
%
%
\author{Albin Forsberg\inst{1}\orcidID{0009-0009-5780-4492} \and
Leonardo Horn Iwaya\inst{1}\textsuperscript{(\Letter)}\orcidID{0000-0001-9005-0543}}
\authorrunning{A. Forsberg and L. H. Iwaya}
%
\institute{Privacy and Security Research Group, Department of Mathematics and Computer Science, Karlstad University, Universitetsgatan 2, 651 88, Karlstad, Sweden\\
\email{leonardo.iwaya@kau.se}}
\maketitle  
\begin{abstract}
Mobile health applications (mHealth apps), particularly in the health and fitness category, have experienced an increase in popularity due to their convenience and availability. However, this widespread adoption raises concerns regarding the security of the user's data. In this study, we investigate the security vulnerabilities of ten top-ranked Android health and fitness apps, a set that accounts for 237 million downloads. We performed several static and dynamic security analyses using tools such as the Mobile Security Framework (MobSF) and Android emulators. We also checked the server's security levels with Qualys SSL, which allowed us to gain insights into the security posture of the servers communicating with the mHealth fitness apps. Our findings revealed many vulnerabilities, such as insecure coding, hardcoded sensitive information, over-privileged permissions, misconfiguration, and excessive communication with third-party domains. For instance, some apps store their database API key directly in the code while also exposing their database URL. We found insecure encryption methods in six apps, such as using AES with ECB mode. Two apps communicated with an alarming number of approximately 230 domains each, and a third app with over 100 domains, exacerbating privacy linkability threats. The study underscores the importance of continuous security assessments of top-ranked mHealth fitness apps to better understand the threat landscape and inform app developers.
\end{abstract}
\section{Introduction}
Mobile health applications (mHealth apps) have become increasingly popular due to their availability, convenience, and ability to provide accessible health-related solutions to users across multiple devices~\cite{momen2020measuring,iwaya2020security}. While many subcategories of mHealth apps exist, some are more popular and widely used than others.
Health and fitness are the most popular categories and have seen a massive increase in usage, especially between January 2019 and January 2020, when the downloads of the most popular health and fitness apps worldwide almost doubled from 8.84 million in 2019 to 16.28 million in 2020~\cite{statista-health-apps3}.

However, previous studies examining the security posture of mHealth apps have found worrying results~\cite{Art3,Art5}, ranging from insecure communication, weak encryption, poor access control, and unfair or unavailable privacy policies. 
Such findings stress the need for continuous research to assess these threats' extent and impact on mHealth apps' data security.
For such reasons, this paper aims to assess the security vulnerabilities of top-ranked mHealth apps for health and fitness. Hence, to guide this study, we developed the following research questions (RQs):
\begin{itemize}
    \item \textbf{RQ1:} \textit{What are the security risks in top-ranked health and fitness apps?} \textbf{Objectives:} to identify the security risks specific to health and fitness apps and to conduct penetration tests to evaluate the extent to which these risks exist in current health and fitness apps.
    \item \textbf{RQ2:} \textit{How can these identified risks be mitigated?} \textbf{Objective:} to propose recommendations for improving the security of health and fitness apps.
\end{itemize}

Briefly, the study's methodology involved three key stages: (i) app selection, (ii) static and dynamic security analyses, and (iii) responsible disclosure. The app selection process employed a Google Play Scraper~\footnote{Google Play Scraper (\url{https://github.com/JoMingyu/google-play-scraper})} to filter apps available in four English-speaking countries, with a minimum rating of 4 stars and over 1 million downloads, thus focusing only on top-ranked apps. For the security analysis, various tools were used, including the Mobile Security Framework (MobSF~\footnote{MobSF is an open-source automated security testing tool for mobile apps (\url{https://github.com/MobSF/Mobile-Security-Framework-MobSF}).}) for static and dynamic analysis, Qualys SSL for server-side assessment. A thorough manual analysis of the tools' outputs was conducted in the analysis process. Lastly, the responsible disclosure process involved compiling all findings into comprehensive reports and communicating them to the respective app companies.

This study contributes with valuable insights into the security posture of some of the most popular and top-ranked mHealth fitness apps, empirically assessing them and identifying the prevalence of vulnerabilities within the apps.
Furthermore, this study proposes recommendations and mitigation tactics for the identified vulnerabilities, aiming to assist developers and stakeholders in improving the security of their apps.
Parts of this work first appeared in the thesis of Forsberg (2024)~\cite{forsberg2024penetration}, in which readers can find a more extensive description of the research.

\section{Related Work}
Many mHealth apps today collect and handle sensitive user data without employing appropriate security measures~\cite{Art1,Art2,Art3}. Studies have shown that many mHealth apps rely on vulnerable communication, transmitting data without encryption and sending sensitive user information over insecure channels~\cite{Art1,Art2}. In addition, a majority of analyzed mHealth apps have been found to transmit sensitive data without encryption over the Internet or use weak encryption and hashing methods such as Electronic Code Book (ECB) cipher mode and Message-Digest 5 (MD5) algorithm~\cite{Art7}. In fact, for many years, studies have highlighted issues such as the misuse of cryptographic Application Programming Interfaces (APIs) and insecure Initialization Vectors (IVs)~\cite{Art10}. Recent research further emphasizes that insecure encryption practices remain prevalent in mHealth apps, with findings revealing the widespread use of vulnerable IVs and ciphers~\cite{Art3}.

For instance, an empirical study~\cite{Art3} conducted penetration tests on 27 mHealth apps, revealing numerous security vulnerabilities. These included information disclosure, weak access control, and excessive data permissions, raising severe concerns about user data privacy. Similarly, another study~\cite{Art5} highlighted issues such as excessive permissions and unauthorized data transmission to third parties, with instances of location tracking without user consent.
Even earlier studies, such as~\cite{Art4}, have highlighted significant shortcomings in mHealth app security, finding that out of 154 mHealth apps studied, health data encryption was rarely provided~\cite{Art4}.

Contributions from other researchers evaluating the privacy of mHealth Apps have come to similar conclusions and findings.
For instance,~\cite{Art5} found that some applications requested permissions that stretched beyond the intended scope. These permissions included access to Bluetooth and the microphone without any apparent need. The conclusion drawn from the situation was that an ad library used Bluetooth permission to track users' location~\cite{Art5}.
Notably, the same study also found that 35\% (7/20)~\cite{Art5} of apps transmitted the postal address or geolocation to vendors or third parties, whereas three apps did so insecurely over HTTP.

In addition to examining client-side security measures, studies have conducted comprehensive analyses of Secure Sockets Layer (SSL) web servers using the Qualys SSL Labs tool~\cite{Art5,Art3}. This tool performs a series of tests on specified web servers, evaluating the validity and trustworthiness of certificates and inspecting server-side SSL configurations. According to the findings of~\cite{Art5}, out of 117 HTTPS connections established with third-party servers, 108 servers had a grade of C or higher, while six servers received an F and three received a T grade. Moreover, out of 11 HTTPS connections made to the app vendors, five were rated A\footnote{Further details on the Qualys SSL Server Rating Guide can be found at \url{https://github.com/ssllabs/research/wiki/SSL-Server-Rating-Guide}}, three were rated B, one was rated C, and two were rated T~\cite{Art5}.

Given that, this study focuses on the top-ranked mHealth apps for health and fitness. Our particular focus on top-ranked apps is currently under-researched since many publications attempt to assess a high number of apps (e.g.,~\cite{Art1,Art2,Art10}), thereby analyzing apps that are not actually extensively used in the real world. For this reason, we prioritize the security assessment of apps that can indeed impact millions of users. We also seek to verify and compare findings with prior work, providing further insights into the security status of mHealth apps.

\section{Methods}
\subsection{App Selection Process}
We focus on selecting top-ranked mHealth fitness apps across four English-speaking countries: Australia, Canada, the United States, and the United Kingdom. The Google Play Scraper tool was used to find the top 30 apps from each country under the search terms \textit{``health and fitness''} (searches finalized on February 15, 2024). Only the apps presented in all four countries were retained for consistency.
Two main criteria were established to select top-ranked apps: (i) apps with at least 1M downloads and (ii) with a star rating of 4 or higher.
These criteria allowed us to narrow it down to 15 apps. However, after attempting to locate these apps' APK files, only 10 top-ranked apps remained to be analyzed.
We purposely selected only the top-ranked apps for this study since they have a broad user base that may be affected.

\subsection{Security Analysis Process}
Our methodology is similar to that of~\cite{Art3}, yet focuses more on security than privacy. The analysis begins with a semi-automated \textbf{static analysis} based on MobSF, comprising several manual inspection steps summarized in Table~\ref{tab:MobSFStatAnTests}. This involves evaluating permissions, code, hardcoded secrets, manifest configurations, and domain assessments to verify potential security vulnerabilities.

\begin{table}[ht]
    \centering
    \small
    \caption{MobSF Static Analysis Main Tests.}
    \begin{tabular}{|lp{0.74\textwidth}|}
        \hline
        \textbf{Tests} & \textbf{Descriptions}\\
        \hline \hline
        App's permissions & Evaluate if the requested permissions align with the app's intended functions, detecting potential overprivileged permissions. \\
        Abused Permissions & Flag permissions that are commonly abused by known malware. \\
        Network Security & Analyze network behavior for vulnerabilities like insecure protocols or unencrypted data transmission. \\
        Certificate analysis & Checks digital certificates for issues like expiry or trust, highlighting security weaknesses. \\
        Manifest file analysis & Review the app's manifest for misconfigurations or vulnerabilities. \\
        Code analysis & Conducts static analysis to identify security vulnerabilities or coding errors. \\
        Trackers & Identifies third-party trackers collecting user data without consent. \\
        Hardcoded Secrets & Search for sensitive information directly embedded in the code, exposing security risks. \\
        \hline
    \end{tabular}
\label{tab:MobSFStatAnTests}
\end{table}

It is worth noting that MobSF is prone to produce many false positives for certain vulnerabilities~\cite{Art3,Art5} as it may not always have access to the complete context of an application's behavior or environment. Therefore, the results obtained from the static analysis output by MobSF were verified manually by the first author and discussed among the team. The following steps were taken:
\begin{itemize}
  \item \emph{Inspecting Code:} Review flagged code to rule out false positives in MobSF's code analysis for insecure number generators, ciphers, and cipher modes.
  \item \emph{Evaluating Permissions:} Assess whether permissions labeled as ``dangerous'' are truly necessary for the app's intended functionality.
  \item \emph{Assessing Hardcoded Secrets:} Review the hardcoded sensitive information found in the apps' source code, such as API keys, passwords, or cryptographic keys, present potential security risks.
 \item \emph{Assessing Manifest Misconfigurations:} Evaluate the severity of misconfigurations in the Android manifest, such as allowing an application to be installed on vulnerable unpatched Android versions.
\end{itemize}

Following that, the \textbf{dynamic analysis} in MobSF captures runtime information such as network traffic, system calls, log messages, and API calls. The output of the dynamic analysis includes detailed reports and logs that provide insights into the apps's behavior, enabling us to inspect:
\begin{itemize}
    \item \emph{Logcat Logs:} Logcat logs are system logs that include runtime information, error messages, warnings, and debugging information generated by the Android operating system and the application itself.
    \item \emph{HTTP(s) Logs:} HTTP(s) logs capture the network traffic generated by the apps, such as requests and responses exchanged with remote servers.
    \item \emph{App Data:} Apps store data on the mobile phone, creating folders, files, and databases, which can be inspected to determine if personal data is securely stored.
\end{itemize}

The apps were thoroughly used for approximately 15 minutes each, which was more than enough to access, interact, and input data to all the activities and available features. Fake emails and user accounts were created for testing the apps, enabling us to access and input data, upload images, schedule workouts, edit entered information, and explore all features. We stress that the apps were used only for their intended purposes, e.g., creating profiles, adding goals, scheduling workouts, changing user info, etc. We did not perform any interactions that could tamper with the app and its server-side infrastructure, e.g., interfere with or inject malicious inputs or network traffic.

All the data captured in the dynamic analysis was carefully inspected by the first author, and the team reviewed the findings. 
Logcat logs were also examined to identify whether they reveal any sensitive information about the user's app usage patterns, activities, or web traffic. It is important to note that the Logcat logs generated on the device are accessible to other apps running concurrently, which can lead to the exposure of sensitive information~\cite{HackingAndroid}.
HTTP(S) logs were examined to determine if the app's traffic was transmitted securely (i.e., check for any personal data sent over unencrypted channels) and to create a list of communicating servers.
The app data were analyzed using SQLite Browser, and each stored database query was manually checked for privacy risks and user information.

Based on the captured network traffic, all domains communicating with each app are \textbf{assessed with Qualys SSL}, determining their security levels concerning HTTPS protocol configurations. The analysis provides an overall rating of the web server's security (A+, A, B, C, D, E, F, T) and a score and potential weaknesses for its certificate, protocol support, key exchange, and cipher strength. It is important to note that only the root domains were tested, and subdomains were not individually assessed.

\subsection{Responsible Disclosure Process}
After completing the vulnerability assessment, individual reports outlining the findings for each app were compiled. These reports were emailed to the respective companies and developers on April 9, 2024, utilizing the contact information on the Google Play Store platform or websites. The identified issues were communicated to the companies/developers 60 days before the study was made public~\cite{forsberg2024penetration}, allowing time for them to mitigate the vulnerabilities. Notwithstanding, all the results from this research are published de-identified, without disclosing the names of the apps or companies. We argue that this study aims to understand the prevalence of security issues, which can be done without singling out and identifying app companies/developers.

\subsection{Ethical Considerations}
This project followed internal regulations and received ethical approval from the Ethical Advisor at Karlstad University (registration number HNT 2023/795).
Furthermore, we adhere to ethical hacking guidelines for security research to try to find and mitigate security and privacy vulnerabilities in the apps~\cite{Ethhacking2}. Before publishing our findings, identified vulnerabilities were reported to the respective app companies or developers, allowing sufficient time for remediation efforts. 
It is also worth reiterating that during the tests, we never tampered with the application infrastructure (e.g., injecting network traffic or trying to trigger SQL vulnerabilities, etc.), and we did not collect personal data from any other users.

\section{Results}
Ten top-ranked apps met all the criteria from the initial set of 30 apps across the four English-speaking countries.
Cumulatively, these 10 apps have been downloaded over 237 million times, underscoring the relevance of studying them specifically. Even if they constitute a small sample, negative impacts can be far-reaching.

\subsection{Hardcoded Secrets}
Hardcoding sensitive data such as API keys or cryptographic materials in the app's source code poses significant security risks. Such hardcoded secrets are easily extracted through reverse engineering, and attackers may gain unauthorized access to backend systems or compromise user data, making it an increased security risk~\cite{owasp-mobile-top10-m7,owasp-mobile-top10-m8}.

We found several instances of hardcoded secrets within the code of the tested mHealth apps in the static analysis. Table~\ref{tab:hardcoded_secrets} summarizes the findings related to hardcoded secrets. While some hardcoded secrets might not pose an immediate security threat~\cite{owasp-mobile-top10-m8}, such as a database URL, their collective prevalence is worth mentioning. This study found that eight out of ten apps had hardcoded their Firebase URL (8/10 apps) and the corresponding API key (10/10 apps). However, Firebase API keys are not used for authorization and may not directly lead to unauthorized access~\cite{firebase_api_keys}.

\begin{table}[ht]
    \centering
    \small
    \caption{Summary of Hardcoded Secrets.}
    \label{tab:hardcoded_secrets}
    \begin{tabular}{|p{0.1\linewidth}|p{0.45\linewidth}|p{0.42\linewidth}|}
    \hline
    \textbf{Categ.} & \textbf{Findings} & \textbf{Risks} \\ \hline \hline
    API Keys & -- All ten apps in the study had hardcoded API keys. 
    & -- Hardcoding API keys increase the risk of exposure and unauthorized access to sensitive data or functionalities. \\ \hline
    Database URLs & -- Eight hardcoded database connection URLs were identified within the source code. & 
    -- Carries no immediate risk by itself.
    \\ \hline
    Client Secrets and Tokens & -- Six instances of hardcoded client secrets and tokens used for authentication with providers such as Facebook and other third parties were found. & -- Hardcoding authentication credentials increase the risk of account takeover and unauthorized access to user data. \\ \hline
    \end{tabular}
\end{table}

From these findings, the Client Secrets and Tokens shown in Table~\ref{tab:hardcoded_secrets} are the most concerning since Meta has specified that access tokens should never be hardcoded into client-side code, seeing that a decompiled app can give full access to your app secret~\cite{meta_api_keys}. Therefore, developers should consider the security of client secrets and tokens exposed in the app's code.

\subsection{Code Analysis} \label{sec:codeanalysis}
Examining mobile application code for vulnerabilities is essential for upholding the security and reliability of mHealth apps. Malicious actors can access and browse application binaries with tools such as MobSF, simplifying the reverse engineering process. Hence, vulnerabilities such as weak pseudorandom number generators (PRNGs) or insecure encryption configurations become easy targets for exploitation~\cite{Art3,owasp-mobile-top10-m8}.
The findings from the investigated high-risk flags supplied by MobSF during the static analysis are summarized in Table~\ref{tab:high-risk-issues}.

\begin{table}[ht]
    \centering
    \small
    \caption{Insecure Code in Apps according to MobSF static analysis.}
    \label{tab:high-risk-issues}
    \begin{tabular}{|l|p{0.9\linewidth}|}
        \hline
        \textbf{Apps} & \textbf{High Risk Issues}\\
        \hline \hline
        App1 & App uses the encryption mode CBC with PKCS5/PKCS7 padding. This configuration is vulnerable to padding oracle attacks. \\
        \hline
        App2 & 1. Debug configuration enabled. Production builds must not be debuggable. 2. Remote Web debugging is enabled. \\
        \hline
        App3 & App uses the encryption mode CBC with PKCS5/PKCS7 padding. This configuration is vulnerable to padding oracle attacks. \\
        \hline
        App4 & Uses ECB mode for encryption.  \\
        \hline
        App5 & Debug configuration enabled. Production builds must not be debuggable. \\
        \hline
        App7 & 1. App uses the encryption mode CBC with PKCS5/PKCS7 padding. This configuration is vulnerable to padding oracle attacks. 2. Remote Web debugging is enabled. \\
        \hline
        App8 & CBC with PKCS5/PKCS7 padding, ECB mode  \\
        \hline
        App10 & App uses the encryption mode CBC with PKCS5/PKCS7 padding. This configuration is vulnerable to padding oracle attacks. \\
        \hline    
    \end{tabular}
\end{table}

The analysis revealed several instances of insecure encryption configurations, posing significant and urgent security risks. 
Five applications were found to utilize the encryption mode CBC with PKCS5/PKCS7 padding, which is known to be vulnerable to padding oracle attacks under certain conditions.
While using CBC with PKCS5/PKCS7 padding indicates a potential problem, it is worth noting that this is only an indication and may result in false positives. We recommend further investigation by the app developers.
Additional conditions must be met to perform a padding oracle attack, i.e., the application must return distinguishable error messages or responses based on the validity of the padding. If the ciphertext is authenticated with a Message Authentication Code (MAC), the risk of a padding oracle attack can be significantly reduced. 
Therefore, while MobSF's detection of CBC with PKCS5/PKCS7 padding highlights a potential security issue, developers should interpret it cautiously.

\begin{figure}[ht]
  \centering
  \includegraphics[width=0.85\textwidth]{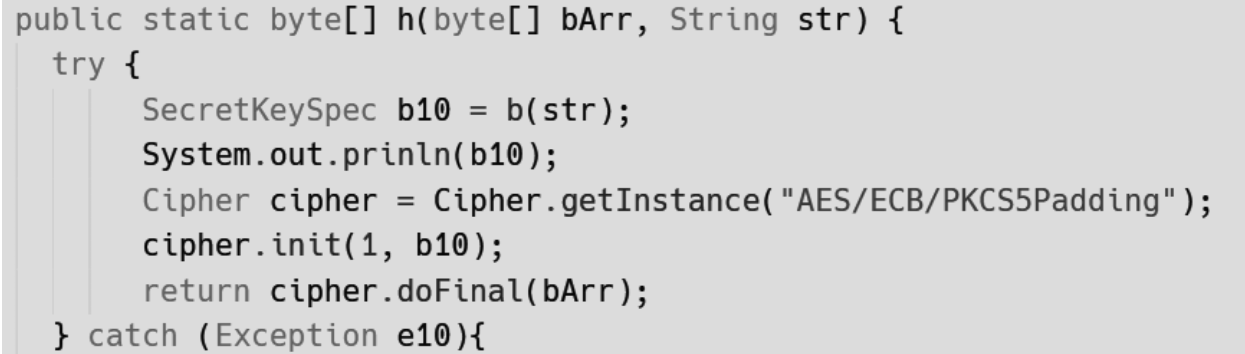}
  \caption{Use of ECB mode by App4.}
  \label{fig:ECB_mode_encryption}
\end{figure}

Additionally, one application was found to use ECB mode for encryption (see Figure~\ref{fig:ECB_mode_encryption}), further increasing security concerns.
Due to its weak encryption, it is not recommended for developers to implement AES/ECB with or without padding~\cite{google-unsafe-encrypt-mode,owasp-ECB-mode} but instead to use AES/GCM/NoPadding as recommended by Google~\cite{google-unsafe-encrypt-mode}.
Although it is hard to determine the exact data that is insecurely encrypted with AES/ECB, such lousy coding practice is alarming and requires immediate attention.
Besides, it is worth mentioning that MobSF's code analysis also flagged several false positives regarding PRNGs, MD5, and SHA1 usage.
These issues were discarded as the app used the algorithms in non-cryptographic use cases (e.g., hashes for integrity checks of downloads).

We also found that three apps had debug configurations enabled, allowing for remote web debugging and production build debugging~\cite{owasp-mobile-top10-m7}. Enabling debug configurations in production builds could expose sensitive information and functionality to potential attackers. Debuggable production builds simplify the task for malicious actors to scrutinize the application behavior and intercept data, potentially resulting in data leaks.

\subsection{Third-party Trackers}
The examined applications were found to communicate with various trackers, as summarized in Table~\ref{tab:trackers-and-apps}. These trackers serve multiple purposes: analytics, advertisement, crash reporting, and identification. Most apps communicate with well-known analytics trackers, including Google Crashlytics, Google Firebase Analytics, and Facebook Analytics, often employed to collect and analyze app usage data, providing developers with insights into user behavior and app performance. Trackers such as Braze (formerly Appboy), AppsFlyer, and Optimizely were also among the most frequently encountered. 

The evaluation of trackers used by the apps revealed several notable findings. Firstly, it was found that at least half of the apps utilize cross-device advertisement networks. 
Sharing information with trackers can raise privacy and security concerns, particularly those used for user profiling and advertisement. By leveraging user data from browsing states, cookies, and browsing history, cross-device ad networks can deliver targeted ads across various devices~\cite{solomos2019talon}.

Furthermore, the analysis identified several apps engaging in user profiling activities using specific trackers such as Tealium, Braze, and Amplitude. These trackers are known for their capabilities in analyzing user behavior and preferences to create detailed user profiles, potentially raising concerns about user privacy and data protection.
Notably, one tracker, Mixpanel, was found to have a concerning privacy breach clause in its policy. According to an analysis by Exodus Privacy~\footnote{Exodus Privacy Trackers List (\url{https://reports.exodus-privacy.eu.org/en/trackers/})}, Mixpanel's policy states that end-users tracked by MixPanel's ``customers'' have no right to delete their personal information~\footnote{Exodus Report on MixPanel: \url{https://reports.exodus-privacy.eu.org/en/trackers/118/}}. This indicates potential privacy implications for users, such as negating data deletion rights, intervenability, and objection to processing.

Additionally, certain trackers such as Branch~\footnote{Exodus Report on Branch: \url{https://reports.exodus-privacy.eu.org/en/trackers/167/}} and Inmobi~\footnote{Exodus Report on Inmobi: \url{https://reports.exodus-privacy.eu.org/trackers/106}} are known to collect device-specific information according to Exodus Privacy, including unique identifiers and other device-related data. This raises concerns about the extent of data collection by these trackers and the potential for user privacy infringement.

\begin{table}[ht]
    \centering
    \small
    \caption{Number of Apps by Tracker.}
    \label{tab:trackers-and-apps}
    \begin{tabular}{|c|l|l|}
        \hline
        \textbf{N. of Apps} & \textbf{Trackers} & \textbf{Used For} \\
        \hline \hline
         7 & Google Firebase Analytics & Analytics\\
         7 & Google CrashLytics & Crash reporting\\
         7 & Facebook Analytics & Analytics\\
         6 & Facebook Login & Identification\\
         5 & Facebook Share & Sharing\\
         4 & Braze (formerly Appboy) & Analytics, Advertisement, Location\\
         3 & Branch & Analytics \\
         3 & Google AdMob & Advertisement\\
         3 & AppsFlyer & Analytics\\
         3 & Optimizely & Analytics\\
         3 & New Relic & Analytics\\
         2 & IAB Open Measurement & Identification, Advertisement\\
         2 & Sentry & Crash reporting\\
        \hline
     \end{tabular}
\end{table}

\subsection{HTTP(S) Traffic Analysis} \label{sec:insecure-communication}
\begin{figure}[ht]
  \centering
  \includegraphics[width=0.9\textwidth]{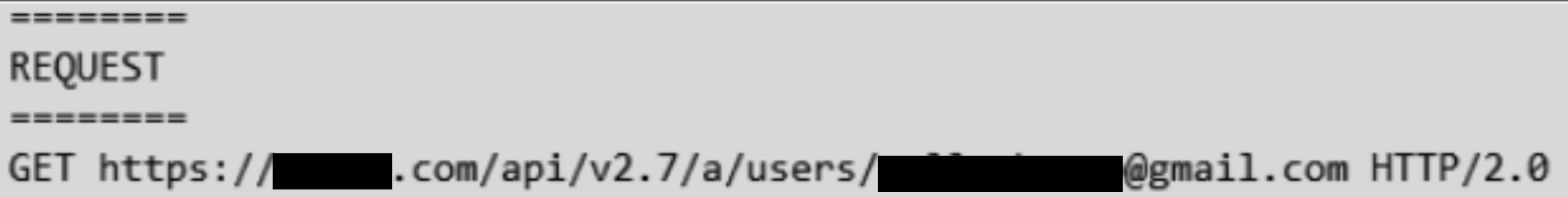}
  \caption{Email used as an identifier in a RESTful URI.}
  \label{fig:GET-email}
\end{figure}

The HTTP(S) traffic analysis yielded reassuring results as it was found that all apps transmitted sensitive user information over encrypted HTTPS connections.
However, two apps were found to use email addresses as identifiers in RESTful URI requests, as depicted in Figure~\ref{fig:GET-email}. Additionally, a third app similarly revealed users' usernames through this method. This practice is generally considered insecure, as using email addresses or usernames in RESTful URI requests exposes user information directly in the request path. Although the contents of HTTPS requests, including paths and query parameters, are encrypted and not exposed during transmission, this information may still be logged on the server side. Consequently, server logs could contain sensitive user information, posing privacy risks if unauthorized individuals access the logs.

Moreover, the analysis revealed that four apps transmitted users' passwords in cleartext POST requests. Even in encrypted POST requests, it is not recommended to keep passwords in plaintext. App developers can already use secure hashing functions on the client side, avoiding transmitting passwords in plaintext and hopefully not storing them in plaintext, which would also be considered a bad practice~\cite{owasp_password_plaintext_storage}.

\subsection{Dangerous Permissions Evaluation}
Android classifies specific permissions as ``dangerous,'' indicating that their misuse or over-privilege could potentially compromise user privacy or device security~\cite{android_permissions_guide}. Over-privileged apps pose risks in terms of both user privacy and security. Unnecessary access to sensitive device features (e.g., external storage, photos, media, contacts list) increases the attack surface for potential exploitation by malicious actors, potentially leading to a breach of user privacy~\cite{raymond2020over}.
The most common permissions used by the apps are shown in Table~\ref{tab:dangerous-permissions}.

Most of the permissions found to be used by the apps could be justified and aligned with the apps' intended purposes. 
However, certain permissions such as \texttt{BLUETOOTH\_SCAN}, \texttt{WRITE\_CONTACTS}, and \texttt{GET\_TASKS} indicate over privilege and raise concerns regarding potential misuse.

The \texttt{GET\_TASKS} permission, although deprecated in API level 21 and no longer enforced, allowed applications to retrieve information about currently and recently running tasks. This permission could potentially enable malicious applications to discover private information about other applications~\cite{chen2023predicting}.

Similarly, the permission of the \texttt{WRITE\_CONTACTS} allows an application to modify the contact data stored on a user's phone. Such permission is unnecessary for the apps' intended use and can be promptly removed.
Furthermore, the \texttt{BLUETOOTH\_SCAN} permission is required for an application to discover and pair nearby Bluetooth devices. While this permission might be needed to pair different activity trackers, the app for mobile phones had no such functionality.

Some permissions collectively raised concerns, suggesting over-privilege even if their individual prevalence could be justified. For instance, four of the apps requested for both \texttt{READ\_EXTERNAL\_STORAGE} and \texttt{WRITE\_EXTERNAL\_STORAGE} while also requesting  \texttt{READ\_MEDIA\_IMAGES} (4/10 apps), \texttt{READ\_MEDIA\_AUDIO} (4/10 apps), and \texttt{READ\_MEDIA\_VIDEO} (3/10 apps) permissions, which should suffice for the app's media-related functionalities. This redundancy suggests a lack of optimal permission management, potentially exposing user data to unnecessary risks.

Excessive location-related permissions across the examined mHealth apps were also of concern. Five apps were found to request both the permissions of \texttt{ACCESS\_FINE\_LOCATION} (5/10 apps) and \texttt{ACCESS\_COARSE\_LOCATION} (4/10 apps), along with additional permissions such as \texttt{ACCESS\_MEDIA\_LOCATION} (3/10 apps). While it is reasonable for fitness-oriented apps to utilize location tracking in the context of, for example, running, the need for multiple location tracking permissions appears excessive and unnecessary. It is also worth noting that the permission \texttt{ACCESS\_MEDIA\_LOCATION} was only justifiable in one of the three apps.

\begin{table}[ht]
    \centering
    \small
    \caption{Most common dangerous permissions used by apps.}
    \label{tab:dangerous-permissions}
    \begin{tabular}{|l|c||l|c|}
        \hline
        \textbf{Dangerous Permissions} & \textbf{\# Apps} & \textbf{Dangerous Permissions} & \textbf{\# Apps}\\
        \hline \hline
        \texttt{POST\_NOTIFICATIONS} & 10 & \texttt{READ\_MEDIA\_VIDEO} & 3 \\
        \texttt{READ\_EXTERNAL\_STORAGE} & 8 & \texttt{ACCESS\_MEDIA\_LOCATION} & 3 \\
        \texttt{WRITE\_EXTERNAL\_STORAGE} & 8 & \texttt{RECORD\_AUDIO} & 2 \\
        \texttt{ACCESS\_FINE\_LOCATION} & 5 & \texttt{BLUETOOTH\_CONNECT} & 2 \\
        \texttt{CAMERA} & 4 & \texttt{BODY\_SENSORS} & 2 \\
        \texttt{READ\_CONTACTS} & 4 & \texttt{GET\_TASKS} & 1 \\
        \texttt{ACCESS\_COARSE\_LOCATION} & 4 & \texttt{GET\_ACCOUNTS} & 1 \\
        \texttt{READ\_MEDIA\_AUDIO} & 4 & \texttt{AUTHENTICATE\_ACCOUNTS} & 1 \\
        \texttt{READ\_MEDIA\_IMAGES} & 4 & \texttt{READ\_MEDIA\_VISUAL\_USER\_SELECTED} & 1 \\
        \texttt{READ\_CALENDAR} & 4 & \texttt{WRITE\_CALENDAR} & 1 \\
        \texttt{USE\_CREDENTIALS} & 4 & \texttt{BLUETOOTH\_SCAN} & 1 \\
        \texttt{ACTIVITY\_RECOGNITION} & 3 & \texttt{WRITE\_CONTACTS} & 1 \\
        \hline
    \end{tabular}
\end{table}

\subsection{Server-Side Analysis}
During the dynamic analysis, we found that some apps communicated with several hundreds of domains over a short testing period (approx. 10-15 min). Table~\ref{tab:dynamic_domains} presents the count of domains communicated with by each app during the dynamic analysis. 
It is apparent from Table~\ref{tab:dynamic_domains} that there is considerable variation among the apps in terms of the number of domains they interact with. For instance, App3 and App9 communicated with relatively fewer domains (0 and 15, respectively), while App8 and App10 were found to interact with an astonishing number of domains (236 and 240, respectively). Altogether, the ten apps communicated with 404 unique domains, i.e., excluding same-origin subdomains.

This significant variation in domain communication patterns is concerning, as there seems to be no justifiable reason for an application to communicate with such a large number of domains within a short time frame.
This behavior raises privacy and security concerns, as apps interacting extensively with domains exacerbate the risks of data leakage, data linkability, and observability. Further examination of these communications and the related domains is essential to measure the extent of the risks associated with such behavior.

In addition, the security of the domains communicated with by the apps and their TLS/SSL configurations was assessed using Qualys SSL. The summarized results are presented in Table~\ref{tab:domain-score-results}. It is important to note that any domain scoring a B or below should evaluate its configurations to enhance security measures, as there is no reason to keep servers configured with lower security levels. Of particular concern is the prevalence of 2.5\% (10 domains) receiving a score of T, indicating multiple misconfigurations or vulnerabilities. Given the popularity of the tested apps, communication with such potentially vulnerable domains requires more immediate attention to mitigate potential security risks promptly.

\begin{table}[ht]
    \centering
    \small
    \caption{Number of domains communicated with the app during dynamic analysis.}
    \label{tab:dynamic_domains}
    \begin{tabular}{|l|c|c|c|c|c|c|c|c|c|c|}
        \hline
        \textbf{Apps} & App1 & App2 & App3 & App4 & App5 & App6 & App7 & App8 & App9 & App10 \\ \hline
        \textbf{Domains} & 17 & 30 & 0 & 111 & 34 & 18 & 33 & 236 & 15  & 240 \\
        \hline
    \end{tabular}
\end{table}

\begin{table}[ht]
    \centering
    \small
    \caption{Summary of Qualys SSL results from web server analysis ($n=404$)}.
    \label{tab:domain-score-results}
    \begin{tabular}{|l|c|c|c|c|c|}
        \hline
        \textbf{Scores} & A+ & A & B & T & No Res \\ \hline
        \textbf{Domains (\%)} & 86 (21.3\%) & 148 (36.6\%) & 107 (26.5\%) & 10 (2.5\%) & 53 (13.1\%) \\
        \hline
    \end{tabular}
\end{table}

\section{Discussion}
This study sought to investigate security vulnerabilities in top-ranked mHealth fitness apps currently used by millions of users. Ten apps were tested for security, revealing many insecure coding practices, excessive tracking, and permissions. In the following subsections, we discuss our main findings and reflect on prior work in the field. We also provide ten recommendations to app developers, which are summarized in Table~\ref{tab:recommendations}.

\begin{table}[ht]
    \centering
    \small
    \caption{Summary of Recommendations to App Developers.}
    \label{tab:recommendations}
    \begin{tabular}{|p{0.99\linewidth}|}
        \hline
        \textbf{App's Security Testing \& Coding} \\
        \hline \hline
        \textbf{1.} Software companies should allocate sufficient resources and budget for security testing. \\
        \textbf{2.} Software companies should enforce secure coding standards and best practices across development teams to ensure consistency and adherence to security guidelines. \\
        \textbf{3}. App developers should implement automated security testing as part of the continuous integration and deployment pipeline to detect vulnerabilities early in the development lifecycle and prevent them from reaching production environments. \\
        \textbf{4.} App developers should avoid including sensitive user information such as email addresses or usernames in GET requests, RESTful URIs, or parameters. Instead, use non-identifiable attributes, such as an ID number or a hashed value. \\
        \hline \hline
        \textbf{Domains \& Trackers} \\
        \hline \hline
        \textbf{5.} App developers should carefully evaluate the necessity of each domain communicated with by their apps and limit communication to only those essential for the app's functionality. \\
        \textbf{6.} Server domain owners should prioritize the security of their TLS/SSL configurations to ensure the confidentiality and integrity of data transmitted between the app and the server. \\
        \textbf{7.} App developers should carefully consider the implications of using cross-platform advertisement trackers and minimize their usage wherever possible, as well as thoroughly review the privacy policies of third-party trackers used in their apps to ensure compliance with privacy regulations and user rights. \\
        \hline \hline
        \textbf{Misconfigurations \& Permissions} \\
        \hline \hline
        \textbf{8.} App developers should run a static analysis (e.g., using MobSF) before each new update to identify possible security risks. \\
        \textbf{9.} App developers should consider adopting a permissions in-context approach, requesting permissions only when users initiate interactions with corresponding app features. \\
        \textbf{10.} App developers should disable non-essential permissions by default, enabling them only when necessary. \\
        \hline 
    \end{tabular}
\end{table}

\subsection{Insecure Coding in mHealth Apps}
Our static analysis revealed hardcoded secrets and weak cryptographic strategies, such as ECB mode and CBC with PKCS5/PKCS7 padding.
Other researchers have reached similar findings in a previous study~\cite{Art7}, showing that out of 30 apps examined, most employed weak cryptographic methods, ECB mode for data encryption, and decryption~\cite{Art7}. 
Many other researchers have reported weak cryptography and insecure configurations in Android apps over the years~\cite{Art3,Art5,Art10}.
It suggests that developers may not be abiding by the basic principles of secure coding and indicates a potential lack of maintenance of the app's security, raising concerns about the overall resilience of these applications against common threats and attacks.
These are well-known vulnerabilities that are easy to discover with tools such as MobSF, which makes their prevalence in this study particularly concerning.

We believe such problems are linked to a knowledge gap on the developer's side. For instance, previous studies have found that 80\% of mHealth app developers had poor security knowledge and that 85\% of developers had an inconsiderable security budget~\cite{aljedaani2020empirical}.
Given the substantial number of coding vulnerabilities uncovered in this study, including weak cryptographic practices, insecure cipher modes, hardcoded secrets, and app behavior logging, we believe this was also the case for the developers of the apps assessed in this study. Otherwise, it would be hard to explain the existence of these common issues as the mitigation for it is not considered complicated or time-consuming.

\subsection{Excessive Communicating Domains and Third-Party Trackers}
We found it essential to include an analysis of the domains communicating with the apps and third-party trackers discovered in the static and dynamic analyses.
We found that three apps collectively communicated with more than 300 unique domains (i.e., App10 ($n=240$), App8 ($n=236$), App4 ($n=111$)). The sheer volume of communication raises suspicion since we only tested the apps to sufficiently cover all its functionalities and input relevant user data (i.e., approx. 10-15 min).
The lack of justification for such extensive interaction with external servers raises additional concerns about the behavior of these apps.
In contrast, the remaining apps maintained communication with a more reasonable number of domains, ranging from 0 to 34. 
The work of~\cite{Art3} reached similar results, highlighting a small subset of apps making excessive HTTP(S) requests to dozens of servers.
In such cases, even if traffic is de-identified (e.g., using a Universally Unique Identifier (UUID) or Android Advertising IDs (AAIDs)), the volume of communication makes it easier to link data and single out users over time, negatively impacting their privacy.

Furthermore, the prevalence of cross-device tracking raises concerns about how user data is shared and utilized for targeted advertising across various devices~\cite{solomos2019talon}.
This practice not only compromises user privacy by potentially exposing their data and preferences but also makes it harder for users to control their data as it is not fully transparent which trackers gather what information. Moreover, trackers such as Tealium, Braze, and Amplitude, which collect, store, and create comprehensive user profiles, should not be allowed in apps that handle sensitive user data. Their prevalence also raises questions about the transparency and consent mechanisms in place for such data collection activities and the potential for misuse of this data.
The identification of Mixpanel's privacy breach clause, which denies users the right to delete their personal information, shines a light on the possible misuse of user data. This clause raises significant concerns about users' ability to control their data and underscores the need for greater transparency and accountability in tracker policies.

We also evaluated each domain TLS/SSL configuration, showing that 29\% had some misconfiguration resulting in a lower score (B 26.5\%, similar to the findings of~\cite{Art5}. Developers should re-assess such domains to exhibit faulty TLS/SSL configurations, as these misconfigurations not only compromise the security of data transmission but also undermine user trust in the platform's overall security measures.

\subsection{Unnecessary Dangerous Permissions}
Regarding the dangerous permissions used by the apps, this study shows relatively positive results, with only three apps having one permission each that did not reflect the app's intended use, i.e., an average of 0.33 unnecessary dangerous permissions per app.
A previous study~\cite{Art3} discovered that, on average, 4.1 permissions of the examined apps' ($n=27$) were unnecessary.
Developers can adopt several strategies to minimize the number of such unnecessary permissions. Firstly, if permissions are not essential for the app's functionality, developers can simply remove them from its configuration file. Additionally, developers should request permissions ``in context,'' i.e., when the user starts interacting with the feature that requires it~\cite{Art3}.
 
Nonetheless, the prevalence of multiple storage permissions raised some concerns about their relevance to the apps' purposes. For instance, eight apps used the permissions \texttt{READ\_EXTERNAL\_STORAGE} and \texttt{WRITE\_EXTERNAL\_STORAGE} while also requesting for \texttt{READ\_MEDIA\_IMAGES} (4/10 apps),\texttt{ READ\_MEDIA\_AUDIO} (4/10 apps), and \texttt{READ\_MEDIA\_VIDEO} (3/10 apps). According to Android's permissions documentation~\footnote{Additional information about Android permissions: \url{https://developer.android.com/reference/android/Manifest.permission}}, starting in API level 33, the \texttt{READ\_EXTERNAL\_STORAGE} permission has no effect and developers are instead advised to use the other \texttt{READ\_MEDIA} variations mentioned above. Developers can shift their usage of \texttt{READ\_EXTERNAL\_STORAGE} to the \texttt{READ\_MEDIA} variations to align with best practices and ensure compatibility with future Android versions.

Furthermore, three apps were found to request multiple redundant location permissions, i.e., \texttt{ACCESS\_COARSE\_LOCATION}, \texttt{ACCESS\_FINE\_LOCATION}, and \texttt{ACCESS\_MEDIA\_LOCATION}.
These permissions grant apps access to various levels of location information, such as approximate phone location, user's exact coordinates, and location information associated with media files, such as photos or videos, stored on the device. Only one app included a functionality that could utilize the \texttt{ACCESS\_MEDIA\_LOCATION}, but the other two did not.
In most cases, \texttt{ACCESS\_FINE\_LOCATION} should suffice for applications that require precise location data for features such as tracking exercise routes or providing location-based reminders. Using \texttt{ACCESS\_COARSE\_LOCATION} may be appropriate for less precise location needs, such as identifying the user's general area for weather updates or nearby points of interest. However, including multiple location tracking permissions without clear justification indicates over privilege and potential privacy concerns.

\section{Limitations}\label{sec: limitations}
Although this study brings valuable insights, readers should consider some limitations.
Firstly, we limited our scope to 10 top-ranked apps for Android on the Google Play Store, available in English-speaking countries. This selection criteria may not fully represent the diversity of global health and fitness apps, even though the most successful apps tend to be in English and marketed worldwide. Nonetheless, the focus on top-ranked apps helped us focus on apps with an extensive reach and user base.
Secondly, we relied significantly on MobSF for the static and dynamic analysis as it is the most established framework for mobile security testing. However, there are always other tools that could have been considered, such as Privado.ai~\footnote{Privado's Open Source Privacy Code Scanning (\url{https://www.privado.ai/open-source})} or more in-depth analysis with Frida~\footnote{Frida scripting tutorial (\url{https://book.hacktricks.xyz/mobile-pentesting/android-app-pentesting/frida-tutorial})} for runtime monitoring and manipulation.
Thirdly, the server-side analysis relied mainly on Qualys SSL, to assess the TLS/SSL configurations. Although this step was found to be rather time-consuming, researchers can also rely on other tools (e.g., testssl.sh~\footnote{testssl.sh is a free command line tool which checks a server's service on any port for the support of TLS/SSL ciphers, protocols as well as recent cryptographic flaws and more (\url{https://testssl.sh/})}) for even more nuanced results.

\section{Conclusion}
In this study, we investigated the security of top-ranked mobile health and fitness apps for Android. Our study found a substantial number of vulnerabilities related to insecure coding practices (10/10 apps), hardcoding sensitive information (10/10 apps), and using known insecure encryption configurations (5/10 apps). Threats also concerned misconfigurations that allowed apps to be installed on older vulnerable Android versions and over-privileged permissions that are unnecessary for the apps' intended use. Some apps also engage with questionable trackers, expose sensitive data in HTTP(s) traffic, log user actions and POST values, and frequent extensive domain communication with inadequate TLS/SSL configurations.

For future research, we encourage expanding the datasets to cover a broader range of health and fitness apps to provide more comprehensive insights into the prevalence of vulnerabilities and potential threats. Additionally, incorporating other tools, such as Privado's privacy code scanning and Frida scripting, could enhance the depth of analysis, allowing for a more nuanced examination of app behaviors and security measures. Future studies might also want to conduct studies over extended periods (e.g.,~\cite{hatamian2019multilateral,momen2020measuring}) to observe how data sharing and access permissions evolve longitudinally.

\begin{credits}
\subsubsection{Acknowledgments.} This work was supported in part by the Knowledge Foundation of Sweden (KKS), Region Värmland (Grant: RUN/230445), and the European Regional Development Fund (ERDF) (Grant: 20365177) in connection to the DHINO 2 project, and Vinnova (Grant: 2018-03025) via the DigitalWell Arena project. We also thank Prof. Dr. Leonardo Martucci for his early feedback on the research, and Dr. Tobias Pulls for reviewing the findings and suggesting improvements to the manuscript.

\subsubsection{\discintname}
The authors declare that they have no known competing financial interests or personal relationships that could have appeared to influence the work reported in this paper.
\end{credits}

%
%
%
\bibliographystyle{splncs04}
\bibliography{references.bib}
\end{document}